\documentclass[%
 aip,
 amsmath,amssymb,
reprint,%
]{revtex4-1}

\usepackage{graphicx}
\usepackage{dcolumn}
\usepackage{bm}

\usepackage[utf8]{inputenc}
\usepackage[T1]{fontenc}
\usepackage{mathptmx}
\usepackage{etoolbox}
\usepackage{siunitx}
\usepackage{url}
\usepackage{hyperref}
\usepackage{braket}
\hypersetup{colorlinks=true,allcolors=blue}
\usepackage[toc,page]{appendix}
\usepackage[final]{changes} 

\makeatletter
\def\@email#1#2{%
 \endgroup
 \patchcmd{\titleblock@produce}
  {\frontmatter@RRAPformat}
  {\frontmatter@RRAPformat{\produce@RRAP{*#1\href{mailto:#2}{#2}}}\frontmatter@RRAPformat}
  {}{}
}%
\makeatother
\begin{document}

\preprint{AIP/123-QED}

\title{Optimised spectral purity of unfiltered photons via pump and nonlinearity shaping}

\author{Tommaso Faleo}
\email{Tommaso.Faleo@uibk.ac.at}
\affiliation{Institut für Experimentalphysik, Universität Innsbruck, Technikerstr. 25, 6020 Innsbruck, Austria}
\author{Christopher L. Morrison}
\affiliation{Institute of Photonics and Quantum Sciences, School of Engineering and Physical Sciences, Heriot-Watt University, Edinburgh, EH14 4AS, UK}
\author{Roméo Beignon}
\affiliation{Institut für Experimentalphysik, Universität Innsbruck, Technikerstr. 25, 6020 Innsbruck, Austria}
\author{Francesco Graffitti}
\affiliation{Institute of Photonics and Quantum Sciences, School of Engineering and Physical Sciences, Heriot-Watt University, Edinburgh, EH14 4AS, UK}
\author{Vikas Remesh}
\affiliation{Institut für Experimentalphysik, Universität Innsbruck, Technikerstr. 25, 6020 Innsbruck, Austria}
\author{Stefan Frick}
\affiliation{Institut für Experimentalphysik, Universität Innsbruck, Technikerstr. 25, 6020 Innsbruck, Austria}
\author{Alessandro Fedrizzi}
\affiliation{Institute of Photonics and Quantum Sciences, School of Engineering and Physical Sciences, Heriot-Watt University, Edinburgh, EH14 4AS, UK}
\author{Gregor Weihs}
\affiliation{Institut für Experimentalphysik, Universität Innsbruck, Technikerstr. 25, 6020 Innsbruck, Austria}
\author{Robert Keil}
\affiliation{Institut für Experimentalphysik, Universität Innsbruck, Technikerstr. 25, 6020 Innsbruck, Austria}

\date{\today}
\begin{abstract}

Photonic quantum technologies rely on the efficient generation and interference of indistinguishable photons.
Exceptional achievements in this respect have been obtained by domain engineering of quasi-phase-matched parametric down-conversion sources, demonstrating high two-photon interference visibility using only moderate bandpass spectral filtering. 
Here, we optimised the spectral purity and indistinguishability of photons from telecom-wavelength sources by combining Gaussian quasi-phase-matching with Gaussian pump spectral shaping.
Without spectral filtering, we used time-of-flight spectrometry to estimate an upper bound spectral purity of \qty{99.9272(6)}{\percent}, and achieved visibilities of up to \qty{98.5 \pm 0.8}{\percent} in two-photon interference experiments with independent sources.

\end{abstract}

\maketitle


Single-photon sources play a crucial role in the advancement of photonic quantum technologies~\cite{O'Brien2009, Flamini2019, Frick2023}, enabling groundbreaking applications in various fields such as quantum communication and cryptography~\cite{Bennett1993, Gisin2002, Bennett2014, Koji2015, Xu2020}, quantum metrology and sensing~\cite{Slussarenko2017, Lyons2018, Thekkadath2020}, or quantum computing~\cite{Knill2001, Nielsen2004, Walther2005, Raussendorf2001, Bartolucci2023}.
The ability to generate indistinguishable photons at high rates is the key starting point for these applications.
Among many techniques, spontaneous parametric down-conversion (SPDC)~\cite{Burnham1970, Mandel1985} has emerged as a robust and versatile approach for generating heralded single photons~\cite{Kwiat1995, MeyerScott2020, VanDerMeer2020} and entangled photon pairs~\cite{Fedrizzi2007, Meraner2021, Faleo2024}.
It offers room-temperature operation, ease of implementation, and flexibility in generating photon pairs at desired wavelengths.
However, the spectral purity of the generated pairs of photons is often limited by frequency correlations introduced by the energy and momentum conservation in the down-conversion process.
Such a reduced spectral purity directly translates to partial distinguishability between heralded photons from independent sources, which hinders two-photon interference (TPI)~\cite{HOM1987} and poses substantial obstacles for many photonic quantum technologies~\cite{Kok2007}.
These spectral correlations can be mitigated by spectral filtering of the down-converted photons.
However, this introduces additional losses and lowers the heralding efficiency\added{: photon pairs are lost whenever one of the photons' frequency falls outside the filter passband (or the photon gets absorbed because of the finite transmissivity of the filter).}\replaced{ This creates a fundamental trade-off between spectral purity and heralding efficiency that is inherent to any filtering-based approach}{, therefore}, increasing resource demands and reducing scalability~\cite{Branczyk2010, MeyerScott2017}.

To address this challenge more efficiently, research efforts have focused on engineering the joint spectral amplitude (JSA) of SPDC photons by a tailored matching of the down-conversion crystal material, orientation and wavelengths~\cite{Grice2001, Mosley2008}\added{, to intrinsically suppress spectral correlations without sacrificing heralding efficiency for purity}.
\added{In this case, the heralding efficiency is limited only by optical losses, detector efficiency, and the implemented focusing conditions~\cite{Bennink2010}.}
Recently, aperiodic poling of quasi-phase-matched down-conversion crystals~\cite{Tambasco2016, Graffitti2017, graffitti2018independent, Wong2019, JWPan2020, Pickston2021} and pump spectral shaping~\cite{Wong2019} have demonstrated significant \replaced{reduction}{suppression} of spectral correlations.
Building on these advances, we present a telecom-wavelength (\qty{1550}{\nano\meter}) type-II SPDC source based on potassium titanyl phosphate (KTP)\replaced{, that, for the first time, combines aperiodically poled crystals with Gaussian nonlinearity profile and Gaussian pump spectral shaping to minimize spectral correlations.}{crystals, in a Sagnac loop configuration~\cite{Fedrizzi2007,Faleo2024}, that combines apodized nonlinear crystals with tailored pump spectral shaping.}
\replaced{While in previous works either Gaussian nonlinearity profiles were engineered with uncontrolled pump spectra~\cite{Pickston2021} or the pump bandwidth was contolled with fixed, steep-edged spectral filtering~\cite{Wong2019}, our approach employs a programmable spatial light modulator for high-fidelity Gaussian amplitude shaping of the pump, paired with optimized domain-engineered crystals~\cite{Graffitti2017}. This combined optimization yields a predicted spectral purity of $\qty{99.99}{\percent}$}{By introducing an optimized Gaussian nonlinearity profile and employing a tailored Gaussian pump spectrum, we predict a spectral purity of $\qty{99.99}{\percent}$.}
Using time-of-flight spectrometry (TOFS) to reconstruct the joint spectral intensity (JSI), we measured an upper bound spectral purity of \qty{99.9000(2)}{\percent} and inferred a counting noise-free upper bound spectral purity of \qty{99.9272(6)}{\percent}.
Moreover, we demonstrate two-photon interference visibilities of up to \qty{98.5 \pm 0.8}{\percent} between photons from separate emissions \added{of independent sources}, without spectral filtering.
As two-photon interference visibility is affected by factors other than spectral purity, this measurement establishes a lower bound for the spectral purity.
The implemented sources feature a symmetric heralding efficiency of \qty{46.5(1)}{\percent} and a brightness of \qty{4.2}{\kilo\hertz} per \unit{\milli\watt} of pump power.



SPDC is a nonlinear process where a higher-energy pump photon of frequency $\omega_{\mathrm{p}}$ is down-converted into two lower-energy photons of frequencies $\omega_{\mathrm{s}}$ and $\omega_{\mathrm{i}}$, known as the signal and idler photons, respectively.
This process enables efficient production of heralded single photons, where the presence of one photon can be inferred by detecting its partner.
Moreover, using a pulsed laser, photon pairs can be generated at precisely defined time intervals.

We consider here collinear type-II SPDC in KTP crystals characterised by a nonlinear susceptibility $\chi^{(2)}$~\cite{Grice1997}.
The JSA defines the spectral properties of the down-converted photon pairs through two key components: The pump envelope function (PEF), which reflects the spectral properties of the pump beam and energy conservation in the SPDC process; and the phase-matching function (PMF), which arises from the phase-matching condition, a consequence of momentum conservation in the nonlinear crystal, and depends on the optical properties of the medium (see Appendix~\ref{appendix:SPDC} for details).

\begin{figure}
    \centering
    \includegraphics[width=\columnwidth]{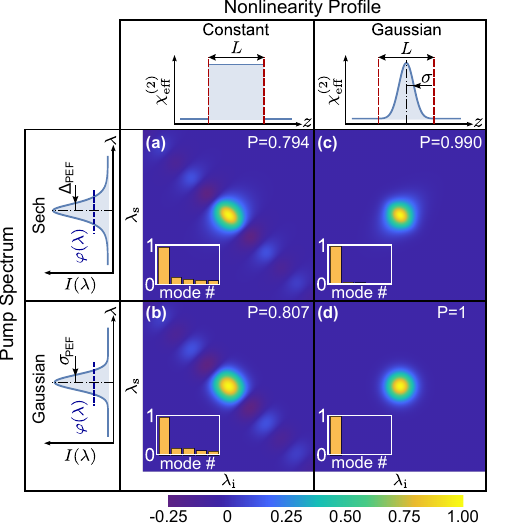}
    \caption{Joint spectral amplitude optimisation under group velocity matching conditions. Panels show the results of different combinations of nonlinearity profiles (top) and pump spectra (left): \textbf{(a)} constant QPM nonlinearity with hyperbolic secant pump, \textbf{(b)} constant QPM nonlinearity with Gaussian-shaped pump, \textbf{(c)} Gaussian-engineered nonlinearity with hyperbolic secant pump, \textbf{(d)} Gaussian-engineered nonlinearity with Gaussian-shaped pump.
    On the top, dashed red lines indicate the \replaced{physical facets of the crystal along the propagation axis $z$, defining the total crystal length $L$}{nonlinear crystal boundaries}, and $\sigma$ denotes the width of the Gaussian nonlinearity profile.
    The pump intensity spectra on the left assume flat spectral phases $\varphi(\lambda)$ (dashed blue lines) with widths ($\Delta_{\mathrm{PEF}}$ for hyperbolic secant, $\sigma_{\mathrm{PEF}}$ for Gaussian) chosen to maximise the spectral purity $P$ of each case.
    The reported values show the ideal purity calculated through Schmidt decomposition and the weights of the first five Schmidt modes.
    }
    \label{fig:purity_optimisation}
\end{figure}

A critical figure of merit for quantum photonic applications is the spectral purity $P$ of the JSA.
High spectral purity ($\approx 1$) enables heralded photons from different sources to interfere with a high visibility in TPI measurements, while lower purity reduces the interference visibility~\cite{Branczyk2017}.
The spectral purity can be quantified through Schmidt decomposition of the JSA as $P=1/K$, where $K$ is the Schmidt number of modes~\cite{Mosley2008,Zielnicki2018}.
It has been predicted that an ideal JSA---where spectral correlations between signal and idler photons are entirely suppressed---can be obtained in crystals characterised by a Gaussian PMF when using a suitably matched Gaussian PEF~\cite{Quesada2018} and group velocity matching~\cite{Keller1997, Grice2001}. 

Figure~\ref{fig:purity_optimisation} shows how this ideal condition can be achieved by shaping the nonlinearity profile and pump spectrum.
The standard phase-matching approach for collinear SPDC, known as quasi-phase matching (QPM), is reported in the first column.
This employs periodic poling of the sign of the nonlinearity $\chi^{(2)}$ to obtain a constant effective nonlinearity $\chi^{(2)}_{\text{eff}}$ along the whole length of the crystal~\cite{Hum2007, Graffitti2018}.
The PMF is linked to the nonlinear profile via a (windowed) Fourier transform, see Appendix~\ref{appendix:SPDC}.
Therefore, the constant nonlinear profile of QPM crystals produces a typical sinc-like shaped PMF, which introduces side lobes in the JSA (see Fig.~\ref{fig:purity_optimisation}(a) and (b)) leading to frequency correlations in the down-converted photons and reducing the spectral purity~\cite{Tambasco2016}.
Moreover, typical picosecond and femtosecond pulsed lasers exhibit hyperbolic secant spectra.
This further limits the maximum purity of the generated photons with respect to a Gaussian spectrum, as shown in the comparison between Fig.~\ref{fig:purity_optimisation}(a) and (b).
Efficient suppression of the spectral correlations arising from the PMF can be achieved with a Gaussian nonlinearity profile that smoothly varies within the length $L$ of the crystal, i.e., with width $\sigma \ll L$, as demonstrated in Fig.~\ref{fig:purity_optimisation}(c).
Such a nonlinearity profile can be experimentally engineered through aperiodic poling of the sign of $\chi^{(2)}$ to best approximate the target joint signal and idler field corresponding to a Gaussian PMF~\cite{Tambasco2016, Graffitti2017}.
Starting from this nonlinearity profile, the ideal purity is obtained when employing a Gaussian pump spectrum with a suitable spectral width $\sigma_{\text{PEF}}$ that renders the JSA circular symmetric~\cite{Graffitti2018}, as highlighted in Fig.~\ref{fig:purity_optimisation}(d).
This can be achieved, for example, by tailoring the spectrum of the pump beam using a pulse shaper~\cite{Monmayrant2010}.

\begin{figure}
    \centering
    \includegraphics[width=\columnwidth]{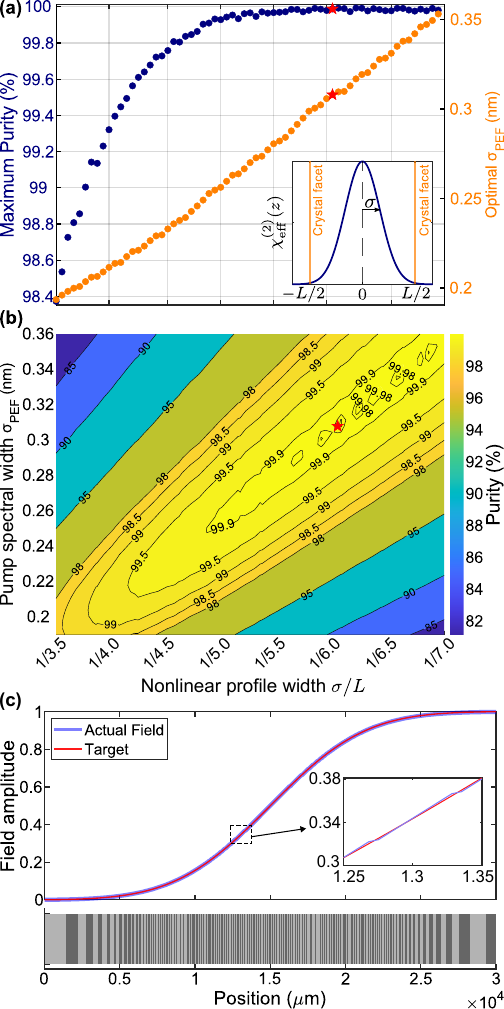}
    \caption{Spectral purity analysis.
    \textbf{(a)} Maximum purity (blue dots) and optimal $\sigma_{\mathrm{PEF}}$ (orange dots) as a function of the nonlinear profile width $\sigma/L$ normalised to the crystal length. Red stars mark the working point $\sigma/L = 1/6.04$. The inset displays the Gaussian target function $g_{\mathrm{ideal}}$ at this optimal width. 
    \textbf{(b)} Purity as functions of normalised width $\sigma/L$ and pump spectral width $\sigma_{\mathrm{PEF}}$. The red star marks the optimal working point ($\sigma = L/6.04$, $\sigma_{\mathrm{PEF}} = \qty{0.308}{\nano\meter})$.
    \textbf{(c)} Comparison between the target field amplitude at $\sigma/L=1/6.04$ and the actual field obtained using the domain engineering in the lower panel. Light and dark grey regions have inverted nonlinearity signs \added{and the width of the smallest domain used is \qty{23.1}{\mu\meter}, corresponding to one coherence length}. The inset demonstrates the accuracy of target field tracking.
    }  
    \label{fig:purity_vs_sigma}
\end{figure}


\begin{figure*}
    \centering
    \includegraphics[width=\textwidth]{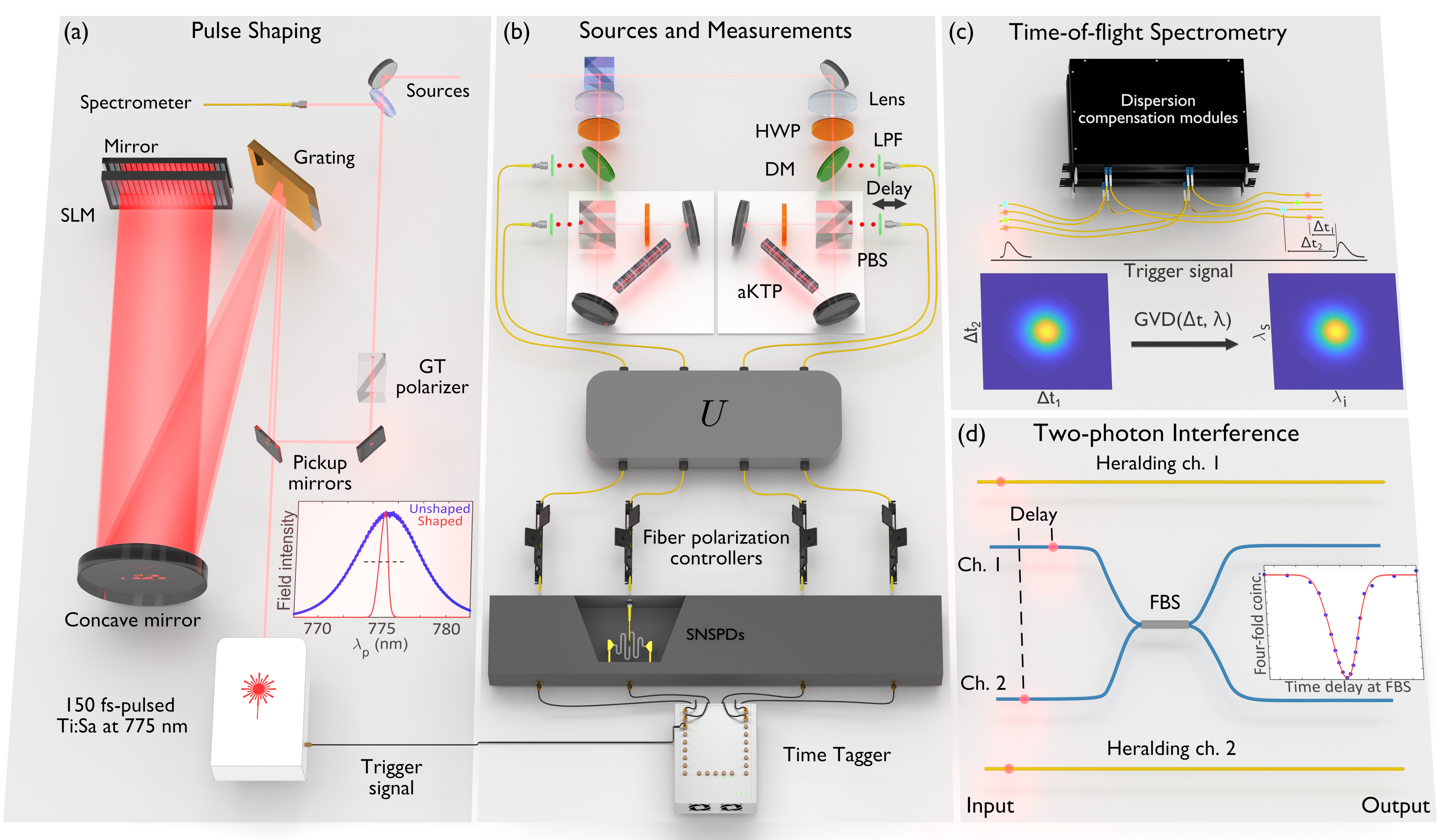}
    \caption{Experimental scheme. \textbf{(a)} A femtosecond (fs) pulsed laser operating at a repetition rate of \qty{76}{\mega\hertz} is spectrally shaped by a spatial light modulator (SLM) positioned at the Fourier plane of a folded 4$f$ pulse shaper.
    A concave mirror ($f=\qty{500}{\milli\meter}$) maps different wavelengths of the first-order diffracted beam to specific pixels of the SLM.
    The beam is reflected with a slight vertical tilt, offsetting the incoming and outgoing beams.
    Amplitude shaping is achieved by combining the polarisation rotation imposed on each wavelength by the SLM with the polarisation rejection carried out by the Glan-Taylor (GT) polariser.
    A beam sampler sends part of the beam to the spectrometer, while the rest is sent to the down-conversion sources.
    The inset shows an example of pump spectrum before and after pulse shaping (the dashed line is the spectral phase for transform-limited pulses)
    \textbf{(b)} The beam is split to the two sources and focused with lenses of focal length $f_{\mathrm{lens}}=\qty{500}{\milli\meter}$.
    Each source consists of a custom apodised-KTP crystal inside a Sagnac loop~\cite{Fedrizzi2007}.
    The four down-converted photons at telecom wavelengths are separated from the pump beam using dichroic mirrors (DM) and lowpass filters (LPF).
    Depending on the measurements to perform, photons undergo a certain single-particle transformation $U$, and single-count and coincidence events between the photons and the laser trigger signal are detected using superconducting nanowire single-photon detectors (SNSPDs) via a time-tagging system.
    \textbf{(c)} In the time-of-flight spectrometry measurements, the signal and idler photons from each source are sent to commercial plug-and-play dispersion compensation modules that introduce different relative time delays depending on the frequencies of the photons.
    A 2D histogram is acquired by recording three-fold coincidences between the trigger signal and the down-converted photons for several combinations of time delays between these channels. The JSI in the inset is reconstructed from this histogram by knowing the chromatic dispersion of the dispersion compensation modules.  
    \textbf{(d)} In the heralded two-photon interference measurements, four-fold coincidences are recorded when interfering the two idler photons from each source at a polarisation-maintaining fibre beam splitter (FBS) while varying their relative time delay at the FBS.}
    \label{fig:experimental setup}
\end{figure*}

The optimisation of the PMF and PEF requires proper selection of the spatial width $\sigma$ of the Gaussian nonlinearity profile with respect to the total crystal length $L$~\cite{Pickston2021}, and of the corresponding pump spectral width $\sigma_{\mathrm{PEF}}$.
While $\sigma/L\ll1$ is required to obtain high purity, this condition leads to a lower total effective nonlinearity averaged across the length of the crystal for a fixed $L$, reducing the brightness of the down-conversion source at a fixed pump power.
To evaluate an optimal trade-off, we \replaced{simulated the domain structures}{engineered the domains} for several $\sigma/L$ ratios and calculated the variation of purity for different values of $\sigma_{\mathrm{PEF}}$.
The results are presented in Fig.~\ref{fig:purity_vs_sigma}, where the obtained values are specific to KTP crystals with $L=\qty{30}{\milli\meter}$ and for degenerate down-conversion to \qty{1550}{\nano\meter}.
The simulation in Fig.~\ref{fig:purity_vs_sigma}(a) shows how changing this $\sigma/L$ ratio leads to higher maximum purities at lower ratios, reaching a plateau of $P=\qty{99.98(1)}{\percent}$ at $\sigma/L \lesssim 1/6$, where only the tails of the Gaussian nonlinear profile are cut at the crystal's facets.
To obtain these optimal spectral purities, the width of the Gaussian pump spectrum $\sigma_{\mathrm{PEF}}$ (measured as intensity-rms) must be matched, as illustrated in Fig.~\ref{fig:purity_vs_sigma}(b).
Smaller $\sigma/L$ ratios produce broader PMFs and require a wider pump spectrum to achieve optimal spectral purity, as shown in Fig.~\ref{fig:purity_vs_sigma}(a).
In order to maximise the spectral purity without exceedingly sacrificing source brightness, we chose to \replaced{fabricate}{work with} crystals \replaced{only at the selected ratio}{engineered for a nonlinear profile with} $\sigma/L = 1/6.04$.
\added{We note that this trade-off between spectral purity and source brightness is distinct from the trade-off introduced by spectral filtering. Here, the absence of filters leaves the heralding efficiency unaffected by the choice of $\sigma/L$.}
According to the simulation, this nonlinear profile delivers a maximum purity of $\mathrm{P}=\qty{99.99}{\percent}$ when using a PEF spectral width of $\sigma_{\mathrm{PEF}} = \qty{0.308}{\nano\meter}$.
Figure~\ref{fig:purity_vs_sigma}(c) shows the result of the domain engineering for this working point.
These results were obtained by using the coherence length as the minimum domain width.
Given the excellent purity achieved in the simulations, this approach was chosen over sub-coherence-length domain engineering, which is more beneficial when shorter crystals and pulse durations are used~\cite{Graffitti2017, Frick} (see Appendix~\ref{appendix:purity comparison} for a comparison between these approaches).


In Fig.~\ref{fig:experimental setup} we present the experimental implementation of our source, together with the pump spectral shaping and characterisation scheme.
The target Gaussian spectrum is shaped out of the broadband spectrum of a tunable Ti:Sapphire laser (pulse duration \qty{150}{\femto\second}) centred at \qty{775}{\nano\meter} via a folded 4$f$ pulse shaper equipped with a programmable spatial light modulator (SLM) positioned at its Fourier plane~\cite{Monmayrant2010, Kappe2025}, as depicted in Fig.~\ref{fig:experimental setup}(a) (see details in Appendix~\ref{sec:methods}).

\begin{figure}
    \centering
    \includegraphics[width=\columnwidth]{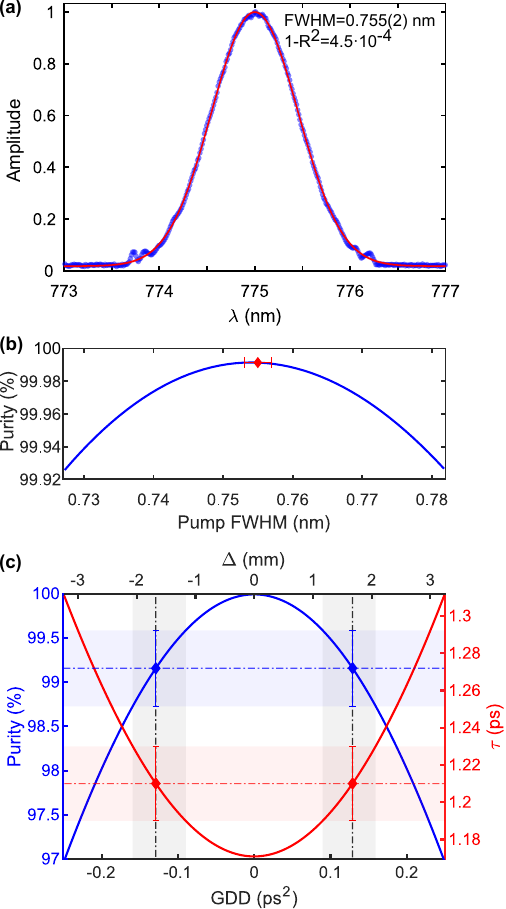}
    \caption{Results Gaussian spectral shaping. \textbf{(a)} Shaped pump amplitude spectrum ($\sqrt{I}$) and corresponding Gaussian fit (red line) with $\sigma_{\mathrm{PEF}}=\qty{0.321(1)}{\nano\meter}$.
    FWHM is defined as full width at half maximum of the intensity.
    \textbf{(b)} Expected purity as a function of the pump bandwidth. The red marker indicates the purity expected from the FWHM in panel (a), with a reduction from the maximum purity of $\approx 10^{-4}\%$. \textbf{(c)} Calculated purity and pulse duration (solid lines), at $\sigma_{\mathrm{PEF}}=\qty{0.321}{\nano\meter}$, as a function of group delay dispersion from the pump spectral shaping. The panel uses dual x-axes: displacement $\Delta$ from optimal grating-to-mirror distance $f$ (upper scale) and corresponding GDD (lower scale).
    The intersection between the measured pulse duration of \qty{1.21(2)}{\pico\second} (red dashed line and shaded uncertainty region) and the simulated curve defines the two GDD intervals (dashed black lines and grey-shaded areas) and the corresponding $\Delta$.
    The intersection of the GDD intervals with the simulated purity establishes the maximum purity of \qty{99.16(43)}{\percent}.
    }
    \label{fig:results1}
\end{figure}

The shaped pump beam is then focused into our custom-made apodised KTP (aKTP) crystals (see Fig.~\ref{fig:experimental setup}(b)). 
We selected the focusing conditions for the pump beam and the collection optics to achieve an optimal trade-off between source brightness and symmetric heralding efficiency (see details in Appendix~\ref{sec:methods}).
Under these conditions, the optimal PEF width shifts to $\sigma_{\mathrm{PEF}}\qty{\approx 0.320}{\nano\meter}$, and full width at half maximum (FWHM) $\qty{\approx 0.754}{\nano\meter}$, slightly larger than the value in Fig.~\ref{fig:purity_vs_sigma}(a).

Figure~\ref{fig:results1}(a) shows the spectral shaping achieved at the output of the pulse shaper to match this ideal PEF.
The R-squared value for a Gaussian fit exceeds $\qty{99.9}{\percent}$, demonstrating the precision of the pulse shaper in accomplishing a Gaussian PEF.
Importantly, the purity exhibits a quadratic dependence on pump spectral width near the optimum, making the system tolerant to small deviations in FWHM.
Minor discrepancies from the ideal value, such as the measured $\mathrm{FWHM}=\qty{0.755}{\nano\meter}$, therefore introduce only negligible purity reductions, as confirmed by the simulation in Fig.~\ref{fig:results1}(b).
By contrast, precise alignment of the pulse shaper is more critical.
Millimetric variations $\Delta$ in the grating-to-concave mirror distance introduce significant group delay dispersion (GDD) \cite{Kappe2025}, which considerably stretches the pulse from its transform-limited duration.
This results in spectral phase correlations in the JSA, reducing its purity~\cite{Graffitti2018}.
We predict this purity reduction with the simulations in Fig.~\ref{fig:results1}(c), which shows the maximum attainable visibility in the presence of a misalignment $\Delta$ and the related GDD.

We arranged both crystals in a Sagnac loop configuration~\cite{Fedrizzi2007, Meraner2021, Faleo2024}, as depicted in Fig.~\ref{fig:experimental setup}(b).
This configuration enables the generation of polarisation-entangled photon pairs if required.
A polarisation beam splitter (PBS) and a dichroic mirror spatially separate the down-converted photons at the outputs of each loop.
Prior to collection into single-mode fibres, lowpass filters further suppress the background from the pump laser.
Importantly, no additional narrow-band filter was added.
To characterise the two sources, we used the collected photons in TOFS or TPI measurements (symbolised by $U$ here; see also Fig.~\ref{fig:experimental setup}(c)-(d)).
Superconducting nanowire single-photon detectors (SNSPDs) with \qty{\approx 85}{\percent} detection efficiency detect the single-photon events at the output, and a time-tagger records both single and coincidence events.
The total optical loss affecting each down-converted photon, excluding their coupling into the fibres, is estimated to be \qty{\approx 19}{\percent}.
The raw, uncorrected, heralding efficiency of the sources is obtained by measuring the single count rates and coincidence rates over \qty{60}{\second}.
We measured single-count rates of $8.5\,\mathrm{kHz/mW}$ and $9.6\,\mathrm{kHz/mW}$ per mW pump power, and coincidence rates of $4.2\,\mathrm{kHz/mW}$, corresponding to a symmetric heralding efficiency of $\eta_{\mathrm{h}} = \qty{46.5(1)}{\percent}$.


The upper bound spectral purity of the down-converted photons is investigated by reconstructing the JSI with TOFS~\cite{Avenhaus2009}.
To do this, we used commercial infrared C-band dispersion compensation modules, as shown in Fig.~\ref{fig:experimental setup}(c) (see details in Appendix~\ref{sec:methods}).
We obtained the JSA by taking the square root of the JSI, assuming a transform-limited joint distribution~\cite{Wong2019}, and we estimated the spectral purity via Schmidt decomposition.
The calculated purity represents an upper bound limit because the JSI measurements are insensitive to the spectral phase correlations of the JSA. 

\begin{figure}
    \centering
    \includegraphics[width=\columnwidth]{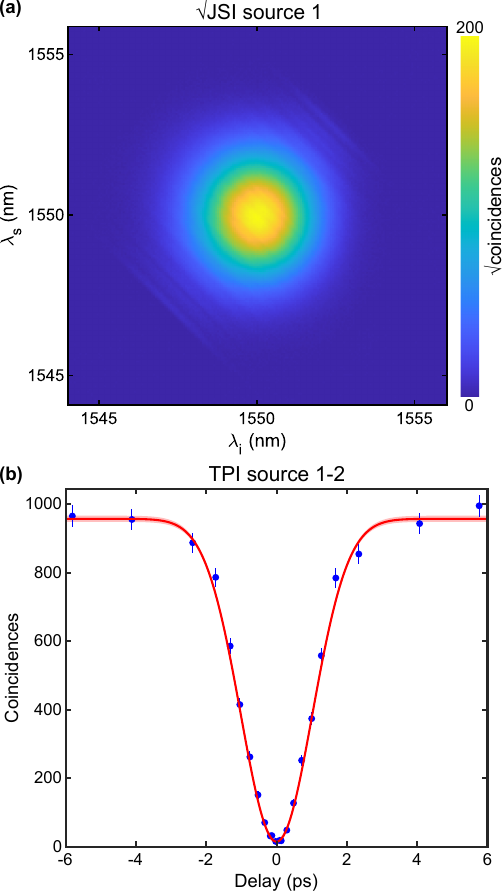}
    \caption{Time-of-flight spectrometry and two-photon interference results. \textbf{(a)} JSA reconstructed from TOFS, with purity of $\qty{99.90}{\percent}$ and estimated maximum purity of \qty{99.9272(6)}{\percent} (see Appendix~\ref{appendix:max purity}).
    \textbf{(b)} Heralded TPI measurement from different sources at a pump power of 5.8 mW, integrating each point over 10 minutes. The red line indicates the fit to the data, and the shaded area is the associated one-sigma uncertainty region. Using only lowpass filters, we obtained a fitted visibility of $\qty{98.5 \pm 0.8}{\percent}$ and a visibility of $\qty{98.5(4)}{\percent}$ extracted from the minimum data and the average of the four maxima.
    }
    \label{fig:results2}
\end{figure}

Figure~\ref{fig:results2}(a) reports the JSA reconstruction of one source, with similar results obtained for the second source.
By reconstructing and performing the Schmidt decomposition on four independent JSAs (see Appendix~\ref{appendix:max purity}), we measured an average spectral purity of $\qty{99.9000(2)}{\percent}$, demonstrating the excellent synergy between nonlinearity engineering and pump spectral shaping in our setup.
A minor reduction of the purity arises from Poissonian noise.
We estimated the maximum purity corrected for this noise contribution to be $\qty{99.9272(6)}{\percent}$ (see Appendix~\ref{appendix:max purity}).
These values align well with the predictions of simulations in Fig.~\ref{fig:purity_vs_sigma} and, to our knowledge, represent the highest upper bound spectral purity achieved in SPDC sources.
Differences between the reconstructed JSAs from both sources reduce the indistinguishability of the heralded photons, decreasing their TPI visibility~\cite{graffitti2018independent}.
Following Ref.~\cite{Branczyk2017}, we employed the Schmidt decomposition on the reconstructed JSAs from both sources and calculated a maximum expectable TPI visibility of \qty{99.84(1)}{\percent}.

To obtain a lower bound for the spectral purity, we performed heralded TPI measurements (see Fig.~\ref{fig:experimental setup}(d)) using the two independent sources~\cite{Ou1999}.
The heralded photons from each source interfere in a balanced polarisation-maintaining fibre beam splitter (FBS) by matching their arrival time at the FBS.
In Fig.~\ref{fig:results2}(b), we present the recorded coincidence counts at different relative time delays between photons.
The dip in coincidences is fitted with a Gaussian function, adopting a bisquare robust fitting method~\cite{O'Brien1992}, resulting in a fitted visibility of $\qty{98.5 \pm 0.8}{\percent}$.
The visibility computed from the minimum and maximum number of coincidences, averaging over the values of the four data points that form the plateau, is $\qty{98.5(4)}{\percent}$.
This represents a significant improvement over previous results with comparable experiments, where visibilities of $\qty{93.9(1.8)}{\percent}$~\cite{Wong2019} and $\qty{95.3(1)}{\percent}$~\cite{Pickston2021} were achieved without spectral filtering.

The discrepancy between the reconstructed JSA purity and the TPI visibility can be attributed to the following factors.
First, contrary to the JSA reconstruction, the TPI measurement is sensitive to unbalanced splitting ratios of the fibre beam splitter, multi-pair production in the down-conversion process, and to any reduction in the photons' indistinguishability due to mismatch of polarisation or arrival time at the beam splitter, and other experimental imperfections.
We estimated the maximum expectable TPI visibility by considering some of these contributions: multipair production; polarisation mismatch of the two photons; and the unbalanced splitting ratio of the fibre beam splitter (see Appendix~\ref{appendix:visibility estimation}).
These imperfections reduce the expected visibility from the ideal value of \qty{99.84(1)}{\percent} to \qty{99.28(7)}{\percent}.

Second, noise photons in other frequency bands can result in unexpected coincidence events, reducing the TPI visibility.
In previous works, gentle filtering around the down-converted photons' wavelength was introduced to suppress this contribution~\cite{Wong2019,Pickston2021}. In our experiment, however, we could not observe any significant improvement in the TPI visibility by adding $8\,\mathrm{nm}$ FWHM bandpass filters to the interfering photons (see Appendix~\ref{appendix:bandpass filter}), which suggests that no such noise photons are relevant here.

Third, as hinted before, spectral correlations hidden in spectral phases of the joint distribution that are not accounted for in the TOFS measurements can lead to reduced TPI visibility.
To investigate this, we used an intensity autocorrelator to measure the pulse duration of the spectrally shaped pump beam, obtaining a value of \qty{1.21(2)}{\pico\second} (see Appendix~\ref{appendix:autocorrelation trace}), which exceeds the expected transform-limited pulse duration of \qty{1.17}{\pico\second} corresponding to the spectrum presented in Fig.~\ref{fig:results1}(a).
This implies that our pulse shaping system introduces a non-negligible dispersion (equivalently, displacements $\Delta \neq 0$), stretching the laser pulses from \qty{1.17}{\pico\second} to \qty{1.21(2)}{\pico\second} and amounting to a GDD of $0.13^{+0.03}_{-0.04} \, \mathrm{ps^2}$.
Considering this, we find that the achievable maximum purity is \qty{99.16(43)}{\percent}, as shown in Fig.~\ref{fig:results1}(c).
This purity reduction results in a further decrease in TPI visibility from the \qty{99.28(7)}{\percent} discussed above to \qty{98.45(43)}{\percent}, which is consistent with the visibility measured in Fig.~\ref{fig:results2}(b).
It is important to note that an intensity autocorrelator only provides a rough estimate of the pulse duration.
A more accurate measurement of the pump pulse characteristics requires full field retrieval techniques, such as frequency-resolved optical grating (FROG)~\cite{Trebino1997, Stibenz2005}.
Knowing the spectral phase profile, a suitable phase compensation approach can be implemented to ensure a constant spectral phase across the pump spectrum and, therefore, obtain transform-limited pulses.
For instance, with the help of an SLM with phase and amplitude modulations, one can impart arbitrary spectral phase profiles to not only correct for distortions, but also to engineer the JSA at will.

With this work, we demonstrated the optimisation of spectral purity in SPDC sources by combining Gaussian nonlinearity engineering and pump spectral shaping.
Using a 4$f$ pulse shaper equipped with a programmable spatial light modulator, we tailored a Gaussian PEF to the Gaussian PMF of custom-made apodised KTP crystals, achieving an upper bound spectral purity of $\geq \qty{99.9}{\percent}$, measured by time-of-flight spectrometry, and a lower-bound spectral purity of \qty{98.5 \pm 0.8}{\percent}, measured by two-photon interference from independent sources, without bandpass spectral filtering.
We provided an in-depth comparison between these measurements by analysing the experimental imperfections of our setup and demonstrating the consistency of the results.
These results significantly advance the current state-of-the-art~\cite{Wong2019, Pickston2021} and, to our knowledge, represent the highest reported unfiltered upper bound spectral purity estimation and two-photon interference visibility from SPDC.
Although characterised by a lower upper bound spectral purity compared to our results, higher interference visibility between independent photon sources has only been reported by using a fully integrated, on-chip spontaneous four-wave mixing platform~\cite{PsiQuantum2025}.
The possibility of working with these high spectral purities without requiring additional spectral filtering enables heralded single-photon sources to achieve the maximum attainable heralding efficiency.
Moreover, we demonstrated two-photon interference between two independent sources, extending beyond the single-source approach of Ref.~\cite{Wong2019} and showcasing the potential for scaling to multiple source scenarios, which will be essential in any photonic quantum technology requiring indistinguishable photons.

As highlighted in Ref.~\cite{Pickston2021}, it is always possible to adjust the input power to match the photon generation rates of brighter periodically-poled sources, with equal contributions from multi-pair production, unless strict restrictions are imposed on the available pump power.
Conversely, it is not possible to increase the spectral purity of spectrally correlated standard QPM sources without reducing the heralding efficiency due to spectral filtering.
Our source therefore operates at the required brightness and optimal spectral purity without sacrificing the heralding efficiency, as it does not require spectral filtering, paving the way for an ideal parametric down-conversion single-photon source. 

The use of a programmable SLM enables fast and flexible spectral shaping of a broadband pulsed laser.
In this work, we took advantage of these features by harnessing an iterative routine optimising the required Gaussian spectrum matching to the Gaussian nonlinearity.
More generally, this versatility can be employed to produce tunable multi-mode frequency-bin entangled states~\cite{Morrison2022, Drago2022, Hurvitz2023, Shukhin2024} for quantum information protocols~\cite{Reimer2019, Lu2020, Folge2024}.
Furthermore, simultaneous phase-amplitude shaping techniques allow full control over the spectral-temporal domain of the pump pulses, enabling access to the temporal-mode degrees of freedom~\cite{Brecht2015, Ansari2018, Graffitti2020, Chiriano2023}, enhancing the resources for high-dimensional quantum information processing.

\section*{Supplementary material}
\noindent See the \hyperlink{supplementary material}{supplementary material} for additional details on: (I) the model of the SPDC process, (II) spectral purity comparison for different domain engineering algorithms, (III) experimental methods, (IV) noise analysis of the reconstructed JSA, (V) two-photon interference visibility estimation, (VI) influence of bandpass filters, and (VII) autocorrelation measurement.

\section*{Acknowledgements}

\noindent This research was funded in part by the Austrian Science Fund (FWF) projects 10.55776/FG5, 10.55776/F71, 10.55776/W1259, 10.55776/COE1, 10.55776/TAI556 and infrastructure funding from FFG (grant no. FO999896024).
C. L. M, F. G., and A.F. acknowledge support by the UK Engineering and Physical Sciences Research Council (grant nos, EP/T001011/1, EP/Z533208/1).
For open access purposes, the authors have applied a CC BY public copyright license to any author-accepted manuscript version arising from this submission.

\section*{Author declarations}

\subsection*{Conflict of interest}
The authors have no conflicts to disclose.

\subsection*{Author contributions}

\textbf{Tommaso Faleo:} Conceptualization; Investigation; Methodology; Formal analysis; Data curation; Visualization; Validation; Software; Project administration; Writing – original draft, Writing – review \& editing.
\textbf{Christopher L. Morrison:} Conceptualization; Methodology; Formal analysis; Software; Writing – review \& editing.
\textbf{Roméo Beignon:} Investigation; Writing – review \& editing.
\textbf{Francesco Graffitti:} Conceptualization; Software; Writing – review \& editing.
\textbf{Vikas Remesh:} Validation; Methodology; Visualization; Writing – review \& editing.
\textbf{Stefan Frick:} Validation; Software; Visualization; Writing – review \& editing.
\textbf{Alessandro Fedrizzi:} Conceptualization; Supervision; Resources; Funding acquisition; Writing – review \& editing.
\textbf{Gregor Weihs:} Conceptualization; Supervision; Resources; Funding acquisition; Writing – review \& editing.
\textbf{Robert Keil:} Conceptualization; Formal analysis; Methodology; Validation; Visualization; Supervision; Funding acquisition; Project administration; Writing – review \& editing. 

\subsection*{Data availability}

The dataset supporting the manuscript is available at \url{https://researchdata.uibk.ac.at//doi/10.48323/p9thx-yak31}.

\section*{References}
\bibliography{references.bib}

\onecolumngrid

\hypertarget{supplementary material}{}
\begin{center}
  \textbf{\Large Supplementary Material}
\end{center}
\appendix

\section{Model of the SPDC process}
\label{appendix:SPDC}
For collinear (without transverse momentum components) type-II SPDC in nonlinear KTP crystals under sufficiently low power conditions, the quantum state of the two down-converted photons can be described as~\cite{Grice1997}
\begin{equation}
    \ket{\psi} = \int \int f(\omega_{\mathrm{s}},\omega_{\mathrm{i}})a_{\mathrm{s}}^{\dagger}(\omega_{\mathrm{s}})a_{\mathrm{i}}^{\dagger}(\omega_{\mathrm{i}}) \ket{0}\mathrm{d}\omega_{\mathrm{s}}\mathrm{d}\omega_{\mathrm{i}},
\end{equation}
where $a_{\mathrm{s}}^{\dagger}(\omega_{\mathrm{s}})$ and $a_{\mathrm{i}}^{\dagger}(\omega_{\mathrm{i}})$ are the creation operators for signal and idler photons at frequencies $\omega_{\mathrm{s}}$ and $\omega_{\mathrm{i}}$, respectively, and
\begin{equation}
    f(\omega_{\mathrm{s}}, \omega_{\mathrm{i}})=\alpha(\omega_{\mathrm{s}}+\omega_{\mathrm{i}})\phi(\omega_{\mathrm{s}}, \omega_{\mathrm{i}})
\end{equation}
is the JSA.
This includes the pump envelope function $\alpha(\omega_{\mathrm{s}}+\omega_{\mathrm{i}})$, expressed in terms of the signal and idler frequencies through energy conservation $\hbar\omega_{\text{p}}=\hbar\omega_{\text{s}}+\hbar\omega_{\text{i}}$, and the phase-matching function $\phi(\omega_{\mathrm{s}},\omega_{\mathrm{i}})$.
The PMF can be determined by the properties of the nonlinear material through the (windowed) Fourier transform~\cite{Grice1997, Tambasco2016, Graffitti2018}
\begin{equation}
    \label{eq:PMF}
    \phi(\omega_{\mathrm{s}},\omega_{\mathrm{i}})=\int_{-L/2}^{L/2} g(z)e^{i \Delta k(\omega_{\mathrm{s}},\omega_{\mathrm{i}}) z}\mathrm{d}z,
\end{equation}
where the integral is performed over the crystal length $L$, which is the direction of propagation.
The first integrand term, $g(z)=\chi^{(2)}(z)/\chi^{(2)}_0$, is the normalized nonlinear susceptibility of the crystal, and $\Delta k(\omega_{\mathrm{s}},\omega_{\mathrm{i}}) = k_{\mathrm{p}}(\omega_{\mathrm{s}}+\omega_{\mathrm{i}}) - k_{\mathrm{s}}(\omega_{\mathrm{s}}) - k_{\mathrm{i}}(\omega_{\mathrm{i}})$ is the mismatch, along the optical axis $z$, of the photons' wavenumber $k_j(\omega_j)=n_j(\omega_j)\omega_j/c$, which depend on the refractive index $n_j(\omega_j)$ of the crystal for each photon $j \in \{ \mathrm{p}, \, \mathrm{s}, \, \mathrm{i} \} $.
The phase-matching condition is satisfied when the mismatch is null.
In quasi-phase matching, the phase-matching condition is satisfied by compensating the non-zero mismatch $\Delta k \neq 0$ with a periodic poling of the crystal with alternate ferroelectric domains of period $\Lambda=2\pi/\Delta k$~\cite{Hum2007}.

For engineered crystals, the nonlinear profile $g(z)$ is designed to approximate the ideal nonlinear profile $g_{\mathrm{ideal}}(z)$ which best approximates the target field amplitude
\begin{equation}
\label{eq:fiel_amplitude}
    A(z, \Delta k) = -i\int_{-L/2}^z g_{\mathrm{ideal}}(z') e^{i \Delta k z'} \mathrm{d}z',
\end{equation}
leading to a Gaussian PMF in Eq.~\eqref{eq:PMF}.
This is implemented through aperiodic poling that tracks the ideal profile, as thoroughly discussed in~\cite{Tambasco2016, Graffitti2017}.

\section{Purity comparison sub-coherence-length}
\label{appendix:purity comparison}

We compare here the performance of the domain engineering implemented in $30 \, \mathrm{mm}$ long aKTP crystals when using coherence-length (\qty{23.1}{\micro\meter}) domain engineering and sub-coherence-length (\qty{2}{\micro\meter}) domain engineering.
To do so, we perform the calculation of the optimal poling of the crystal domains for these two cases, following Ref.~\cite{Graffitti2017} (without using domain-width annealing).
Moreover, we compare these results with a newly developed deterministic sub-coherence-length algorithm~\cite{Frick} that fully takes advantage of the manufacturers' photolithographic resolution capabilities by working on variable domain widths and producing optimal poling structures down to numeric precision.

\begin{figure}
    \centering
    \includegraphics[width=0.55\linewidth]{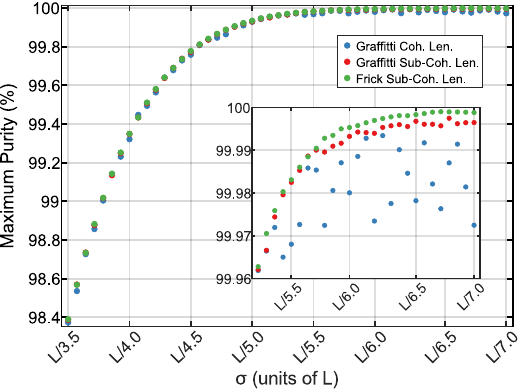}
    \caption{Comparison between domain engineering algorithms. Coherence-length and sub-coherence-length algorithms following Ref.~\cite{Graffitti2017} are labelled with the prefix ``Graffitti". The sub-coherence-length algorithm of Ref~\cite{Frick} is labelled with the prefix ``Frick".}
    \small\textit{Scatter plot comparing maximum spectral purity versus normalised nonlinear profile width for coherence-length and two sub-coherence-length domain engineering algorithms, all converging above 99.98 percent purity for sufficiently narrow profiles.}
    \label{fig:comparison_methods}
\end{figure}

The results for these methods are presented in Fig.~\ref{fig:comparison_methods}. 
All three methods achieve, on average, more than \qty{99.98}{\percent} maximum purity for $\sigma/L<1/6.0$ values, which is more than enough for the purpose of this work.
However, both sub-coherence-length algorithms provide a finer tracking of the ideal field amplitude in Eq.~\eqref{eq:fiel_amplitude}, therefore, showing more consistency in achieving the maximum purity for each chosen $\sigma$ value and providing overall higher maximum purities ($>\qty{99.993}{\percent}$) when $\sigma/L<1/6.0$ compared to the coherence-length case.
In particular, the highest achieved purities are obtained with the algorithm in Ref.~\cite{Frick}.

\section{Experimental Methods}
\label{sec:methods}

\subsection*{Pulse shaper}
Our spectral shaping procedure is based on an intensity-only 4$f$ pulse shaper, in a folded configuration, with a programmable SLM placed in its Fourier plane.
This folded setting is shown in Fig.~\ref{fig:experimental setup}(a).
The diffraction grating (2400 lines/mm) disperses the input beam along the first negative diffraction order towards a concave mirror (CM) with focal length $f=\qty{500}{\milli\meter}$.
The CM collimates the diverging beam across the transverse plane, spatially mapping the dispersed frequency components onto the Fourier plane, which is located at a distance $f$ from the CM, where a reflective mirror is positioned.
The SLM is placed close to the mirror, and the beam propagates through it twice, once before and once after reflection at the mirror.
The CM refocuses the backwards-propagating beam at the grating, which recombines the frequency components at the output.
The 4$f$ condition is fulfilled by setting both the grating-to-CM and CM-to-SLM distances equal to $f$.
To accomplish this condition, we aligned the pulse shaper by minimising the temporal duration of the shaped pump pulses using the feedback of an intensity autocorrelator.

The SLM consists of 128 liquid crystal pixels with a pitch of \qty{98}{\micro\meter}, which corresponds to a spectral resolution of \qty{0.05}{\nano\meter} at the Fourier plane of our setup, measured as the FWHM on a spectrometer.
The voltage applied to each pixel can be individually adjusted to induce a specific polarisation rotation in the light propagating through the liquid crystals.
Combined with an additional polariser after the 4$f$ system, this pixel-by-pixel polarisation control enables precise spectral intensity shaping of the pump beam.
The shaped spectrum is monitored with the spectrometer and fed back to the SLM control software.
The shaping procedure typically converges to the target profile within 20 iterations, corresponding to a few minutes of acquisition time.
The spectral shaping is flexible and stable over time, allowing real-time reconfiguration to compensate for laser spectrum fluctuations if required.

\added{The spectral shaping naturally introduces a reduction in pump power. The optical elements of the folded 4$f$ configuration (diffraction grating, mirrors, SLM) and the Glan-Taylor polarizer at the output have a combined transmission efficiency of approximately \qty{70}{\percent} when the SLM is set to full transmission. The dominant source of power loss, however, is the amplitude shaping itself: since the input spectrum spans \qtyrange[range-units=single,range-phrase=-]{5}{6}{nm} (corresponding to \qty{\approx 150}{fs} pulses), while the target Gaussian spectrum has a FWHM of only \qty{0.755}{nm}, approximately \qty{85}{\percent} of the input spectral power is rejected. These losses do not limit our experiment, as the available laser power (\qty{\approx 2.7}{\watt}) far exceeds the few-mW pump powers required to operate in the low multi-pair emission regime. Furthermore, the shaping loss could be substantially reduced by using a pump laser with a narrower initial bandwidth, e.g., with pulse durations on the order of \qty{1}{ps}.}

\subsection*{Focusing conditions}
The focusing conditions of the pump beam and collection optics play a key role in determining the performance of SPDC sources\added{, as thoroughly discussed in Ref.}~\cite{Bennink2010}.
This is described in terms of the focal parameter $\xi_{\text{j}}=L/k_{\text{j}}w_{\text{j}}$, where $k_{\text{j}}$ and $w_{\text{j}}$ are the wavenumber and waist radius, respectively, for each photon $j \in \{ \mathrm{p}, \, \mathrm{s}, \, \mathrm{i} \} $.
\added{In an unfiltered source design, as presented in this work, the focusing conditions determines the achievable heralding efficiency when optical losses and detector efficiency are disregarded.}
Weak focusing parameters ($\xi_{\text{j}} \ll 1$), which correspond to large waist radii, optimise the heralding efficiency \added{(with 1 as upper limit)} at the expense of the source brightness.
To find an optimal balance between source brightness and symmetric heralding efficiency, we adapted the analysis of Ref.~\cite{Kolenderski2019} for collinear geometry and for the nonlinearity profile of our aKTP crystals to evaluate the source performance.
Taking into account these results and previous studies of periodically-poled KTP crystals~\cite{Bennink2010}, we selected the waist radius of the pump beam at the focal point to be \qty{\approx 66}{\micro\meter} ($\xi_{\text{p}} \approx 0.48$), while the collection optics waist is \qty{\approx 68.5}{\micro\meter} ($\xi_{\text{s}} \approx \xi_{\text{i}} \approx 0.89$).
Compared to weaker focal parameters, these focusing conditions partly compensate for the reduced effective nonlinearity caused by apodisation, while maintaining a high symmetric heralding efficiency (ideally \qty{\approx 85}{\percent} excluding additional optical and detection losses) and practically the same spectral purity reported in Fig.~\ref{fig:purity_vs_sigma}.
Stronger focusing, on the other hand, decreases both the heralding efficiency and the maximum attainable spectral purity.

\subsection*{TOFS setup}
The TOFS technique exploits chromatic group velocity dispersion (GVD) in optical fibres to map photon wavelengths into different arrival times at the detectors~\cite{Avenhaus2009}.
Our commercial infrared C-band dispersion compensation modules are based on dispersion compensation fibres with negative chromatic dispersion of roughly \qty[per-mode = symbol]{-1350}{\pico\second\per\nano\meter}, equivalent to about $\qty{80}{\kilo\meter}$ of standard telecom fibre.
We measured three-fold coincidences between the two down-converted photons and the trigger signal of the laser pulses, used as the time reference, to reconstruct the JSI in the wavelength domain~\cite{Avenhaus2009}.
We recorded three-fold coincidences with \qty{80}{\pico\second} time bins, corresponding to a spectral resolution of roughly \qty{0.06}{\nano\meter}.
The combined time jitter of the system can be calculated by adding in quadrature the independent jitter contributions of the trigger channel and one of the down-converted photon channels.
These contributions include the jitter from one SNSPD channel, two time tagger channels, and the internal photodiode of the pump laser.
We estimated a total time jitter of \qty{\approx 37}{\pico\second}, which is lower than the used time bins and negligibly affects the JSI measurement~\cite{Zielnicki2018}.

\section{Extrapolation of the Poisson noise-free upper bound purity}
\label{appendix:max purity}

\begin{figure}
    \centering
    \includegraphics[width=\linewidth]{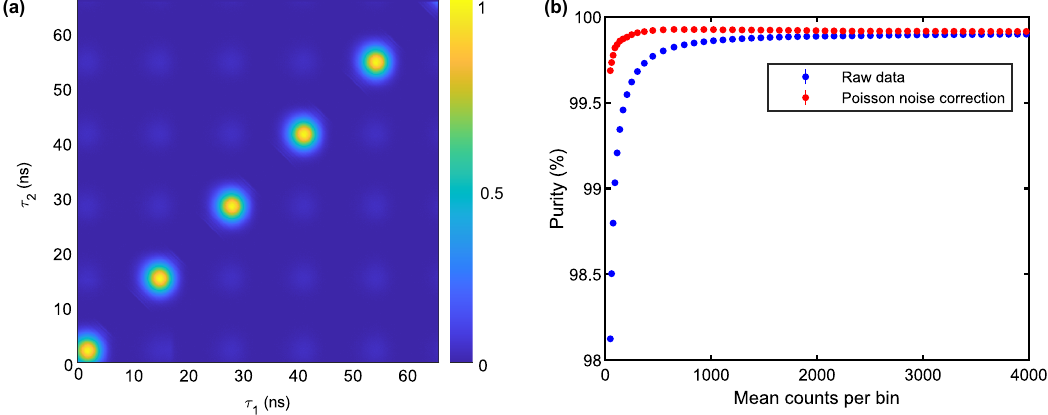}
    \caption{Analysis time-of-flight spectrometry data.
    \textbf{(a)} Three-fold coincidences, as a function of time delays, among laser trigger signal (reference), idler photons, and signal photons. Four JSIs are reconstructed and analysed from the coincidences accumulated in time along the main diagonal.
    \textbf{(b)} Mean purity of the four reconstructed $\sqrt{\mathrm{JSI}}$ as a function of the mean counts per bin, and corresponding purity corrected for the counting noise.
    The maximum corrected purity value is $\qty{99.9272(6)}{\percent}$.
    }
    \small\textit{Two-dimensional coincidence histogram from time-of-flight spectrometry showing periodic joint spectral features along the diagonal, and a plot of measured and Poisson-noise-corrected purity converging as a function of mean counts per bin to a maximum corrected value of 99.9272 percent.}
    \label{fig:max_extracted_purity}
\end{figure}

For our TOFS analysis, we time-tagged three-fold coincidence events among the laser trigger signal, used as the time reference channel, and the detections of idler and signal photons.
Our time-tagging system can acquire two-dimensional histograms with up to $800 \times 800$ time bins of \qty{80}{\pico\second}, as presented in Fig.~\ref{fig:max_extracted_purity}(a).
In this delay map, the coincidence events are accumulated in five features along the main diagonal, spaced by the period of the pulsed laser repetition $\approx \qty{13.15}{\nano\second}$.
We exclude the first feature in the bottom left corner from further analysis, as it is partially cut off by the boundary of the delay map.
The JSI of the source can be obtained by isolating one such feature and converting it to the wavelength domain by knowing the chromatic dispersion applied to the down-converted photons~\cite{Avenhaus2009}, as shown in Fig.~\ref{fig:results2}(a).

The purity resulting from the Schmidt decomposition of the reconstructed JSI is highly dependent on the counting statistics.
For this reason, we acquired three-fold coincidences and sampled the purity of the reconstructed four JSIs as a function of the integration time.
The result is shown in Fig.~\ref{fig:max_extracted_purity}(b).
Initially, the purity rapidly increases as the mean number of coincidences per bin increases in time. 
After roughly 1000 mean counts per bin, the purity increase is slowly converging to a plateau of \qty{\approx 99.900}{\percent}.
Following the procedure in Ref.~\cite{PsiQuantum2025} (see Supplementary information), we also estimated the Poisson noise-free purity by directly correcting the measured values for the fixed offset set by the finite counting statistic.
This results in a maximum purity of \qty{99.9272(6)}{\percent} at $\approx 600$ counts per bin, which decreases and converges towards the raw measured values at higher counts.
This suggests the presence of other purity reduction factors for longer integration times.

\section{Two-photon interference visibility estimation}
\label{appendix:visibility estimation}

Two-photon interference visibility is reduced not only by spectral purity $P<1$, but also by any distinguishability between the photons, which hinders their ability to interfere.
Unbalanced beam splitting ratios provide information about the path of photons entering different input ports of the beam splitter, and different polarisation states make them partially distinguishable.
Moreover, high-visibility two-photon interference relies on the absence of coincidence events, which can arise from multi-pair emissions, i.e., when one or both sources produce more than one down-converted photon pair at a time.  
For these reasons, we precisely analysed the influence of the splitting ratio of our fibre beam splitter, the maximum attainable interference contrast allowed by polarisation overlap at the beam splitter, as well as the contribution of multi-pair events at different pump powers.
Finally, we combined these factors to estimate the maximum expected interference visibility given these three types of experimental imperfections.

\subsection*{Splitting ratio analysis}
We calculated the splitting ratio of our fibre beam splitter following Ref.~\cite{Meany2012}, by reconstructing its $2 \times 2$ unitary transformation $U^{\mathrm{FBS}}$ from measured single-photon counts $C_{i,j}$ between input $i$ and output $j$.
With this method, we can define the ratio
\begin{equation}
    F=\frac{C_{11}C_{22}}{C_{12}C_{21}}=\frac{|U_{11}^{\mathrm{FBS}}U_{22}^{\mathrm{FBS}}|^2}{|U_{12}^{\mathrm{FBS}}U_{21}^{\mathrm{FBS}}|^2},
\end{equation}
which directly connects the measured single-photon counts to the matrix elements of the unitary independently of the input/output losses (set by optical transmission and detector efficiencies).
The transmittance and reflectance of the fibre beam splitter correspond to $\tau=|U_{12}^{\mathrm{FBS}}|^2=|U_{21}^{\mathrm{FBS}}|^2$ and $\rho=|U_{11}^{\mathrm{FBS}}|^2=|U_{22}^{\mathrm{FBS}}|^2$, respectively.
Therefore, they can be obtained as $\tau=1/(1+\sqrt{F})$ and $\rho=\sqrt{F}/(1+\sqrt{F})$.
To evaluate $F$, we measured the counts $C_{i,j}$ by blocking one input at a time and integrating single-photon counts at the two outputs over \qty{60}{\second}.
The results are summarised in Table~\ref{tab:splitting ratio}, where the errors of $\tau$ and $\rho$ are calculated considering a Poisson distribution of $C_{i,j}$.
The maximum achievable visibility with this splitting ratio is $V=\qty{99.990(1)}{\percent}$ \cite{HOM1987}.
We do not multiply this reduction in visibility by the other reduction factors. 
However, the influence of the measured splitting ratio is taken into account when calculating the reduction in visibility caused by multi-pair contributions discussed below.

\begin{table*}
    \centering
    \begin{tabular}{w{c}{3cm}w{c}{3cm}w{c}{3cm}w{c}{3cm}w{c}{3cm}w{c}{3cm}} 
        \hline\\
        $C_{11}$ & $C_{22}$ & $C_{12}$ & $C_{21}$ & $\tau$ & $\rho$ \\[3pt]
        \hline\\
        1386700 & 1577124 & 1593534 & 1333374 & \qty{49.64(2)}{\percent} & \qty{50.36(2)}{\percent}\\[3pt]
        \hline
    \end{tabular}
    \caption{Total number of single-photon counts measured over \qty{60}{\second} for each $C_{i,j}$ and corresponding transmittance and reflectance.}
    \small\textit{Measured single-photon counts at each input-output combination of the fibre beam splitter, yielding a near-balanced splitting ratio of approximately 49.6 percent transmittance and 50.4 percent reflectance.}
    \label{tab:splitting ratio}
\end{table*}

\begin{figure}
    \centering
    \includegraphics[width=\linewidth]{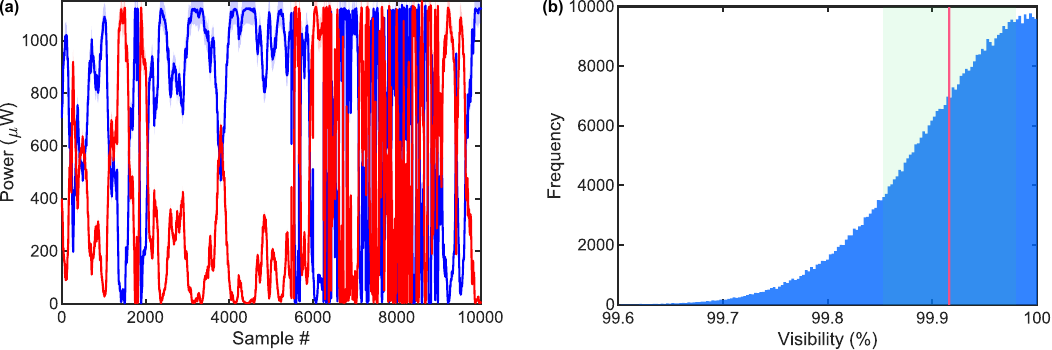}
    \caption{\textbf{(a)} Interference fringes measured at the two outputs of the fibre beam splitter with a linearly polarized \qty{1550}{\nano\meter} laser beam at the inputs. The blue and red shaded areas indicate the one-sigma uncertainty region ($\pm \qty{5}{\percent}$ as declared by the diodes' manufacturer). The power traces are strongly anticorrelated with a Pearson coefficient of $-0.998$ and maximum fringe visibilities of \qty{99.65(2)}{\percent} and \qty{99.55(3)}{\percent} for the blue and red output traces, respectively.
    \textbf{(b)} Frequency histogram of the attainable visibility calculated with Monte Carlo simulations sampling the input powers' discrepancy $\epsilon$ from a Normal distribution $\mathcal{N}(0,0.3)$ (see main text). The red line and the green region are the average value and its one-sigma uncertainty region, corresponding to \qty{99.92(6)}{\percent}.
    }
    \small\textit{Classical interference fringes at the fibre beam splitter outputs showing anticorrelated power traces with visibility exceeding 99.5 percent, and a Monte Carlo histogram of attainable visibility accounting for input power discrepancy yielding a mean of 99.92 percent.}
    \label{fig:polarisation_fringes}
\end{figure}

\subsection*{Polarisation analysis}
Next, we studied an upper bound visibility estimation, taking into account the polarisation of the photons at the input ports of the fibre beam splitter.
When two indistinguishable down-converted photons are collected with their (linear) polarisation aligned to the slow axis of a (balanced) polarisation-maintaining FBS, as depicted in Fig.~\ref{fig:experimental setup}(d), full visibility of two-photon interference occurs when their wave packets are temporally overlapped. 
To assess this, we used a linearly polarised, continuous-wave laser beam with a wavelength of \qty{1550}{\nano\meter} to simulate the propagation of the down-converted photons inside the Sagnacs and their interference at the FBS.
We split the fibre-coupled laser beam using an additional balanced polarisation-maintaining FBS.
By coupling each output of this beam splitter to the (output) heralding channel of each source, the laser propagates from the heralding channel through the Sagnac loop and into the opposite output channel of the source.
There, it is collected in the polarisation-maintaining fibres leading to the FBS used for the TPI in Fig.~\ref{fig:experimental setup}(d).
In each source, the laser undergoes the same polarisation rotations as the down-converted photons due to the optical elements, which determine the alignment of the photons' polarisation with the slow axis of the polarisation-maintaining FBS.
In ideal conditions, the split laser beam undergoes identical polarisation rotations in either source, leading to perfect alignment of the polarisation to the slow axis of the FBS, only potentially picking up different optical phases.
Therefore, the paths through the two sources form a large interferometer.
Consequently, monitoring the power at the outputs of the FBS should reveal an interference pattern as the relative optical phases change.
In such a condition, when the input powers at the FBS are equal and the splitting ratio is 50:50, the interference fringes should exhibit full visibility.
Therefore, any reduction in visibility can be associated with different polarisations at the FBS.

Figure~\ref{fig:polarisation_fringes}(a) shows the time traces of the power measured at the two outputs of the fibre beam splitter when the relative optical phase is altered by introducing an airflow into one loop.
To properly probe the relative phase space, we alternated between a first period with slowly changing phases (0 to 5000 power samples) and a second period with increased airflow and faster power oscillations.
The two power traces are strongly anticorrelated, as expected from the constructive-destructive interference of different outputs, exhibiting high interference visibilities.
To be conservative, we estimated the visibility by considering the global maximum and minimum values of the two traces ($V=(I_{\mathrm{max}}-I_{\mathrm{min}})/(I_{\mathrm{max}}+I_{\mathrm{min}})$), resulting in the maximum visibilities of \qty{99.65(2)}{\percent} and \qty{99.55(3)}{\percent}.
In this estimation of the maximum two-photon interference visibility, we will consider the higher of these two values as an additional contribution to the reduction in visibility.

In order to attribute this reduction solely to differences in polarisation, we correct the raw visibility value by taking into account the imperfect splitting ratio calculated in the previous section, as well as the input power mismatch, which could be maintained within a difference $\epsilon \approx \qty{0.3}{\micro\watt}$.
To evaluate the influence of this mismatch, we performed Monte Carlo simulations of the maximum attainable visibility in the presence of power input discrepancies sampled from a normal distribution centred at $\epsilon=\qty{0}{\uW}$ and with standard deviation $\sigma_\epsilon=\qty{0.3}{\uW}$.
The two input powers are defined in this simulation as $I = I_{\mathrm{avg}} \pm |\epsilon|$, where $I_{\mathrm{avg}}$ is the average power.
The result of these calculations is reported in Figs.~\ref{fig:polarisation_fringes}(b), which shows the frequency histogram of the sampled visibilities, together with the mean value of \qty{99.92(6)}{\percent}.
When including these two effects, the corrected maximum visibility becomes \qty{99.74(7)}{\percent}.

\subsection*{Multi-pair analysis}
The final contribution to this analysis is associated with a reduction in visibility caused by multi-pair creation during the down-conversion process.
This multi-pair production depends on the pair-generation probability per pump pulse, $p$, which is determined by the pump power and the strength of the interaction in the nonlinear crystal.
The probability of down-converting $N$ photon pairs
is $P(N)=(1-p)p^N$~\cite{Jin2015}.
The presence of $N(>1)$-photon states from one source, the other source, or both introduces a non-zero probability of coincidences at the output of the FBS, even with fully indistinguishable photons.
This probability depends on the number $N$ of pairs created by each source, the splitting ratio of the FBS, and the losses in each channel, including the heralding channels.

\begin{figure}
    \centering
    \includegraphics[width=.55\linewidth]{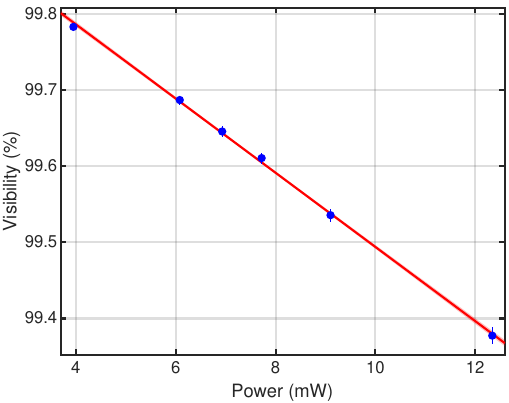}
    \caption{Two-photon interference visibility reduction due to multi-pair emission as a function of the pump power. The linear fit provides a maximum visibility of \qty{99.698(2)}{\percent} at \qty{5.8}{\milli\watt}.
    The red shaded areas indicate the one-sigma fit uncertainty region.}
    \small\textit{Scatter plot with linear fit showing two-photon interference visibility decreasing linearly with pump power due to multi-pair emission, yielding a predicted maximum visibility of 99.698 percent at 5.8 milliwatts.}
    \label{fig:multi_pairs}
\end{figure}

To evaluate the visibility reduction at our pump power (\qty{5.8}{\milli\watt}), we simulated the four-fold coincidence probability for fully distinguishable and indistinguishable photons, taking into account multi-pair creation, FBS splitting ratio, and losses in the setup.
This requires calculating the pair-generation probabilities of the sources, reconstructing the FBS splitting ratio (see the first section of this Appendix) and evaluating the losses (i.e. transmission losses and detector efficiencies).
Following a procedure similar to that in Ref.~\cite{Jin2015}, we connected the outputs of each source directly to the detectors ($U$ in Fig.~\ref{fig:experimental setup}(b) corresponding to the identity matrix here), and we measured single and coincidence events at six pump power values to determine the total losses and pair-generation probability as a function of pump power.
Inserting the FBS requires an additional four fibre matings with an average coupling efficiency of \qty{97(2)}{\percent}, which we included in the total loss.
We used the calculated splitting ratio and losses to reconstruct the overall unitary transformation of the scheme in Fig.~\ref{fig:experimental setup}(d).
To model losses while maintaining unitarity, we extend the four-mode unitary with auxiliary modes that mimic the mechanism of photon loss~\cite{Muenzberg2022}.
Each loss channel is represented by an unbalanced beam splitter with transmittance $\tau$ equal to the channel efficiency, directing lost photons into auxiliary modes with probability $1-\tau$.
We add one auxiliary mode for each heralding channel and one at both the input and output of each FBS channel, resulting in an overall ten-mode unitary.
Given the unitary transformation, for each input state, we calculated the transition probabilities for all possible combinations of outputs resulting in four-fold coincidence events in the non-auxiliary modes.
To account for multi-pair contributions, the input states correspond to different combinations of the number of photon pairs $N$ emitted by the two sources, up to a maximum total of eight input photons, weighted by the associated pair-generation probability.
We performed this calculation for both distinguishable and indistinguishable input photon states, obtaining the visibility as $V=(p_{\mathrm{coinc}}^D-p_{\mathrm{coinc}}^I)/p_{\mathrm{coinc}}^D$, where $p_{\mathrm{coinc}}^D$ and $p_{\mathrm{coinc}}^I$ represent the total computed four-fold coincidence probabilities for the distinguishable and indistinguishable scenarios, respectively.
We implemented these calculations using Monte Carlo simulations to account for the Poisson statistics of single and coincidence events, as well as the uncertainties in the splitting ratios (see Splitting ratio analysis) and optical losses (including the \qty{\pm 3}{\percent} uncertainty on the detector efficiencies declared by the manufacturer).

Figure~\ref{fig:multi_pairs} shows the results of this analysis.
As expected for low pump power and detector dark count rates (\qty{<100}{\hertz} for all detectors)~\cite{Scarani2005, graffitti2018independent}, the visibility exhibits a linear decrease with increasing pump power, yielding a maximum visibility of \qty{99.698(2)}{\percent} at \qty{5.8}{\milli\watt}.

\begin{table*}
    \centering
    \begin{tabular}{w{c}{4cm}w{c}{3cm}w{c}{3cm}w{c}{3cm}w{c}{3cm}} 
        \hline\\
        JSAs independent sources & polarization analysis & Multi-pair analysis & Max visibility & Max visibility (GDD) \\[3pt]
        \hline\\
        \qty{99.84(1)}{\percent} & \qty{99.74(7)}{\percent} & \qty{99.698(2)}{\percent} & \qty{99.28(7)}{\percent} & \qty{98.45(43)}{\percent}\\[3pt]
        \hline
    \end{tabular}
    \caption{Summary of numerical results obtained with the maximum visibility estimation.
    The last two columns respectively differ according to the absence or presence of GDD, as estimated in fig.~4(c) of the main text.}
    \small\textit{Summary of cumulative visibility reduction factors from independent-source JSA overlap, polarisation mismatch, and multi-pair emission, giving a maximum expected two-photon interference visibility of 99.28 percent without and 98.45 percent with group delay dispersion.}
    \label{tab:visibility estimation}
\end{table*}

\subsection*{Summary visibility reduction}

To account for the contributions to visibility reduction calculated in this Appendix, we need to multiply them (independent contributions) by the maximum achievable two-photon interference calculated from the Schmidt decomposition of the JSA from the two sources (see main text).
This results in a maximum two-photon interference visibility that is unaffected by spectral correlations arising from non-null GDD in the pump beam.
However, as discussed in the main text and Appendix~\ref{appendix:autocorrelation trace}, this additional contribution can be estimated from the intensity autocorrelation trace of the pump pulses. 
Table~\ref{tab:visibility estimation} summarises these results, reporting the maximum visibility value for both the GDD-unaffected and -affected scenarios.

\section{Influence of bandpass filters}
\label{appendix:bandpass filter}

\begin{figure}
    \centering
    \includegraphics[width=\linewidth]{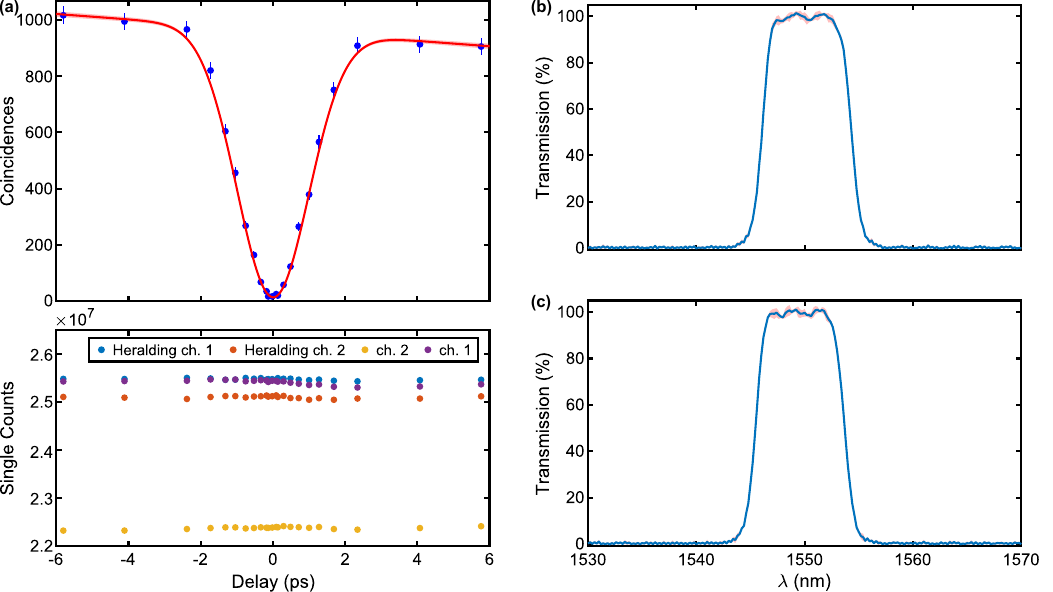}
    \caption{TPI measurement with bandpass filters. \textbf{(a)} Single and coincidence counts measured at a pump power of 5.8 mW with bandpass filters at the collection outputs of the interfering photons. Each point is measured over 10 minutes. The red line indicates the fit to the data, and the shaded area is the one-sigma uncertainty region. We obtained a fitted visibility of $\qty{98.6(0.5)}{\percent}$.
    \textbf{(b)-(c)} Filters' transmission measured as a function of wavelength. The FWHM is \qty{8}{\nano\meter} and the transmission is $T>\qty{98}{\percent}$ over \qty{5}{\nano\meter}. The red shaded area shows the one-sigma uncertainty region.
    }
    \small\textit{Two-photon interference measurement with 8 nm bandpass filters yielding 98.6 percent fitted visibility consistent with the unfiltered result, alongside filter transmission curves showing greater than 98 percent transmission over 5 nm bandwidth.}
    \label{fig:tpi_filters}
\end{figure}

\begin{figure}
    \centering
    \includegraphics[width=0.4\linewidth]{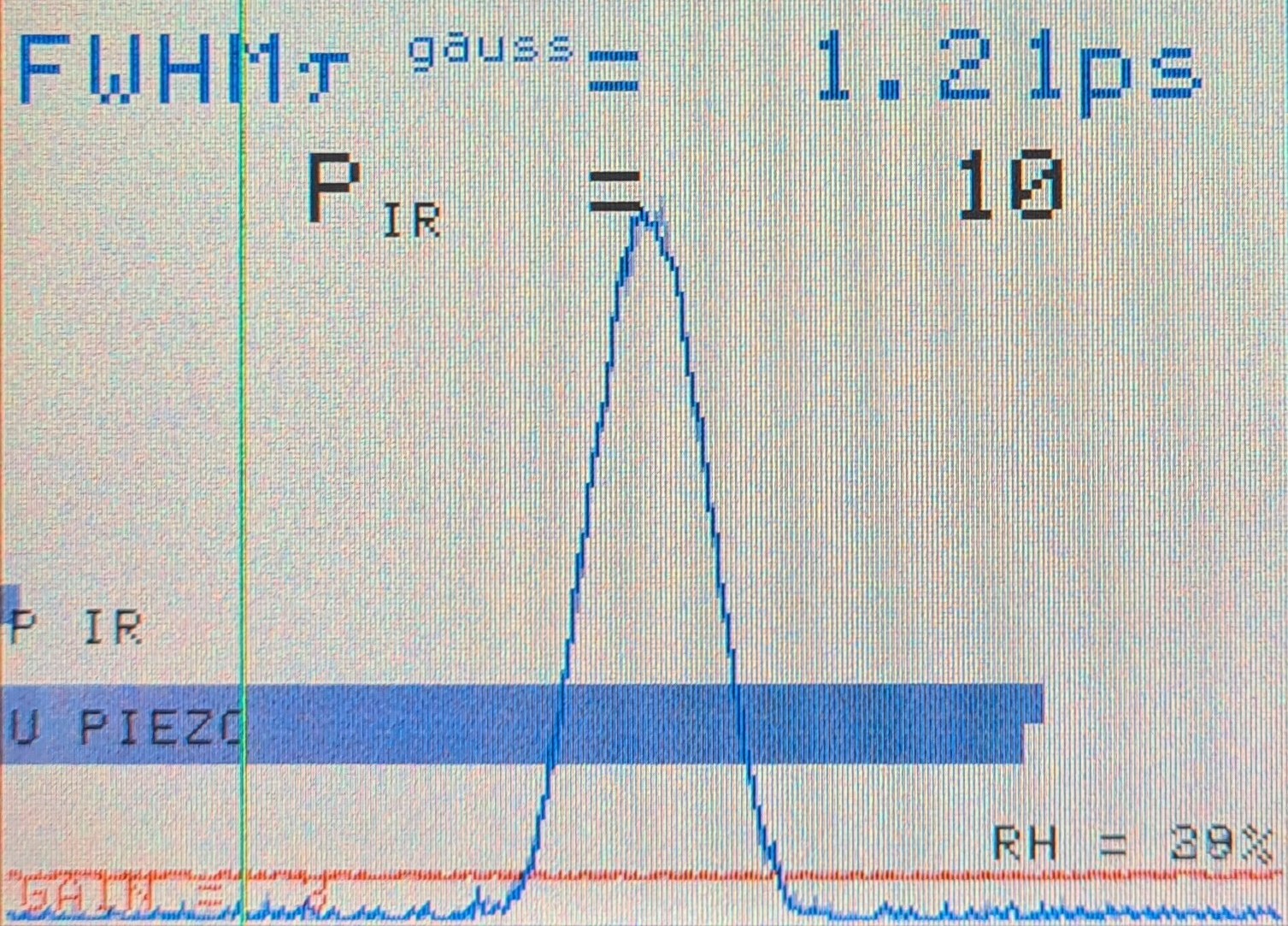}
    \caption{Intensity autocorrelator trace (blue) of the shaped pump beam in Fig.~4(a) of the main text. The measured FWHM assuming a Gaussian temporal shape is \qty{1.21(2)}{\ps}.}
    \small\textit{Intensity autocorrelator trace of the spectrally shaped pump beam showing a measured Gaussian pulse duration of 1.21 picoseconds, slightly exceeding the transform-limited value and indicating residual group delay dispersion from pulse shaper misalignment.}
    \label{fig:autocorrelation trace}
\end{figure}

The visibility of two-photon interference can be influenced by unexpected coincidence events arising from the detection of noise photons in frequency bands different from those of the down-converted photons.
Previous works reported significant increases in two-photon interference visibility by adding moderate bandpass filtering~\cite{Wong2019, Pickston2021}.

For this reason, we performed additional measurements by including bandpass filters in the collection paths of the photons interfering at the fibre beam splitter.
The results of the measurement are presented in Figure~\ref{fig:tpi_filters}.
We fitted the interference pattern in Fig.~\ref{fig:tpi_filters}(a) with a Gaussian function, but we included here a linear term to take into account the overall reduction of coincidences as a function of the delay.
This reduction is independent of the interference and is associated with a slight misalignment when scanning the photons' delay.
This is demonstrated in Fig.~\ref{fig:tpi_filters}(a), which shows that the total number of single counts recorded for channel 1 (the channel used to scan the delay) decreases with increasing delay.
As the number of single counts in the other channels remains roughly constant, this behaviour is most likely due to a slight misalignment caused by the insertion of the bandpass filter in channel 1.
We obtained a visibility of the fitted interference pattern of \qty{98.6(0.5)}{\percent} using bandpass filters with typical transmission curves shown in Fig.~\ref{fig:tpi_filters}(c).
This value is consistent with the interference visibility obtained in the unfiltered case, as reported in the main text (see Fig.~\ref{fig:results2}(b)), suggesting that bandpass filters do not provide additional benefit in our setup.

\section{Autocorrelator trace}
\label{appendix:autocorrelation trace}

Using the pulse shaper requires the 4$f$ distances to be optimally aligned to avoid introducing dispersion in the pump laser.

To tackle this issue, we employed an intensity autocorrelator to provide feedback on the pulse duration and signal intensity, thereby minimising the error from the ideal 4$f$ condition.
Misalignments in the pulse shaper introduce dispersion that increases the pulse duration and reduces the autocorrelation signal (lower peak power decreases the second harmonic generation efficiency).
Therefore, we optimised the 4$f$ setting by maximising the signal and simultaneously minimising the pulse duration.
Figure~\ref{fig:autocorrelation trace} shows the autocorrelator trace obtained after applying this optimisation procedure, with a minimum pulse duration of \qty{1.21(2)}{\pico\second}.
The uncertainty on the reading is considered to be half of the minimum device resolution in this setting, which corresponds to \qty{0.04}{\pico\second}.
The measured trace corresponds to the spectrum presented in Fig.~\ref{fig:results1}(a), which has a transform-limited pulse duration of \qty{1.17}{\pico\second}.
This indicates the presence of a significant temporal stretching of our pump pulses, which suggests a misalignment of roughly $\pm$\qty{1.7}{\milli\meter} in the 4$f$ pulse shaper, as shown in Fig.~\ref{fig:results1}(c).

\end{document}